\documentclass[11pt,fleqn,twoside]{article}
\usepackage{amsfonts,amssymb,latexsym}
\makeatletter
\newcommand{\prava}[1]{\small\it
\begin{flushleft}
Copyright \copyright \ 1999 by  #1
\end{flushleft}}

\newcommand{\name}[1]{\begin{flushleft}
                       \LARGE \bf #1
                       \end{flushleft}\vspace{-3mm}}

\newcommand{\Author}[1]{\begin{flushleft}
                       \it #1 \end{flushleft}}

\newcommand{\Adress}[1]{\begin{flushleft}
                       \it #1 \end{flushleft}}

\newcommand{\Date}[1]{\begin{flushleft}
                      \small  \it #1 \end{flushleft}}

\newcommand{\ehkol}{Author \ name}
\newcommand{\ohkol}{Article \ name}
\renewcommand{\@evenhead}{
\hspace*{-3pt}\raisebox{-15pt}[\headheight][0pt]{\vbox{\hbox to \textwidth 
{\thepage \hfil \ehkol}\vskip4pt \hrule}}}
\renewcommand{\@oddhead}{
\hspace*{-3pt}\raisebox{-15pt}[\headheight][0pt]{\vbox{\hbox to \textwidth 
{\ohkol \hfil \thepage}\vskip4pt\hrule}}}
\renewcommand{\@evenfoot}{}
\renewcommand{\@oddfoot}{}

\newcommand{\be}{\begin{equation}}
\newcommand{\ee}{\end{equation}}
\newcommand{\ba}{\hspace*{-5pt}\begin{array}}
\newcommand{\ea}{\end{array}}

\newcommand{\ds}{\displaystyle}
\makeatother

\begin{document}

\thispagestyle{empty}
\setcounter{page}{263}
\renewcommand{\ehkol}{L.A. Khodarinova and I.A. Prikhodsky}
\renewcommand{\ohkol}{Algebraic Spectral Relations
for Elliptic Quantum Calogero-Moser Problems}

\begin{flushleft}
\footnotesize \sf
Journal of Nonlinear Mathematical Physics \qquad 1999, V.6, N~3,
\pageref{khodarinova-fp}--\pageref{khodarinova-lp}.
\hfill {\sc Letter}
\end{flushleft}
\vspace{-5mm}
\renewcommand{\footnoterule}{}
{\renewcommand{\thefootnote}{}
 \footnote{\prava{L.A. Khodarinova and I.A. Prikhodsky}}}
\name{Algebraic Spectral Relations
for Elliptic Quantum Calogero-Moser Problems}\label{khodarinova-fp}

\Author{L.A. KHODARINOVA~$^{\dag \ddag}$ and  I.A. PRIKHODSKY~$^\ddag$}

\Adress{$\dag$~Wessex Institute of Technology, Ashurst Lodge,
Ashurst, Southampton, SO 047AA UK, \\
~~E-mail: khodarin@wessex.ac.uk \\[1mm]
$\ddag$~Institute of Mechanical Engineering, Russian Academy of Sciences,\\
~~M. Haritonievsky, 4, Centre, Moscow 101830 Russia}

\Date{Received March 29, 1999; Revised April 26, 1999; Accepted
May 17, 1999}

\begin{abstract}
\noindent
Explicit algebraic relations between the quantum integrals
of the elliptic Calogero--Moser quantum problems related to the root systems
${\bf A_2}$ and ${\bf B_2}$ are found.
\end{abstract}

\section{Introduction}

\noindent
The notion of algebraic integrability for the quantum problems
arised as a multidimensional generalisations of the ``f\/inite-gap"
property of the one-dimensional Schr\"odinger operators
(see~\cite{khodarinova:Kr,khodarinova:ChV,khodarinova:VSCh}).

Recall that the Schr\"odinger equation
\[
L\psi=-\Delta\psi+u(x)\psi=E\psi ,\qquad x\in R^n ,
\]
is called {\it integrable} if there exist $n$ commuting dif\/ferential
operators $L_1=L,L_2,\ldots,L_n$ with the constant algebraically independent
highest symbols $P_1(\xi)=\xi^2,P_2(\xi),\ldots,P_n(\xi)$, and
{\it algebraically integrable}
if there exists at least one more dif\/ferential
operator $L_{n+1}$, which commutes with the operators $L_i$, $i=1,\ldots,n$,
and whose highest symbol $P_{n+1}(\xi)$ is constant and takes dif\/ferent
values on the solutions of the algebraic system $P_i(\xi)=c_i$,
$i=1,\ldots,n$, for generic $c_i$.

According to the general result \cite{khodarinova:Kr} in the
algebraically
integrable
case there exists an algebraic relation between the operators $L_i$,
$i=1,\ldots,n+1$:
\[
Q(L_1,L_2,\ldots,L_{n+1}) = 0.
\]

The corresponding eigenvalues $\lambda_i$ of the operators $L_i$ obviously
satisfy the same relation:
\[
Q(\lambda_1,\lambda_2,\ldots,\lambda_{n+1}) = 0.
\]
Sometimes it is more suitable to add more generators from the commutative
ring of quantum integrals: $L_1,L_2,\ldots,L_{n+k}$; in that case we have
more than one relation. We will call these relations {\it spectral}. They
determine a spectral variety of the corresponding Schr\"odinger operator.

The main result of the present paper is the explicit description of the
spectral algebraic relations for the three-particle elliptic
Calogero-Moser problem with the Hamiltonian
\begin{equation}
L=-\frac{\partial ^2}{\partial {x_1^2}}-\frac{\partial ^2}{\partial {x_2^2}}-
\frac{\partial ^2}{\partial {x_3^2}}+4(\wp (x_1-x_2)+\wp (x_2-x_3)+\wp
(x_3-x_1))  \label{khodarinova:1}
\end{equation}
and for the generalised Calogero-Moser problem related to the root system
${\bf B_2}$ (see~\cite{khodarinova:OP}) with the Hamiltonian
\begin{equation}
L=-\frac{\partial ^2}{\partial {x_1^2}}-\frac{\partial ^2}{\partial {x_2^2}}%
+2(\wp (x_1)+\wp (x_2)+2\wp (x_1+x_2)+2\wp (x_1-x_2)).  \label{khodarinova:2}
\end{equation}
Here $\wp $ is the classical Weierstrass elliptic function satisfying the
equation
\[
\wp^{\prime }{}^2-4\wp ^3+g_2\wp +g_3=0.
\]
These operators are known as the simplest multidimentional generalizations
of the classical Lame operator (see~\cite{khodarinova:ChV}).

For the problem~(\ref{khodarinova:1}) reduced to the plane
$x_1 + x_2 + x_3 = 0$ the
explicit equations of the spectral variety have been found
before in~\cite{khodarinova:SV}.
The derivation of ~\cite{khodarinova:SV} is indirect and based on the
idea of the
``isoperiodic deformations"~\cite{khodarinova:GS}. In the present
paper
we give a direct
derivation of the spectral relations for the
problem~(\ref{khodarinova:1})
using
explicit formulae for the additional quantum integrals, which have been
found in~\cite{khodarinova:Kh}. The derived formulae are in a good
agreement with the
formulae written in~\cite{khodarinova:SV}.

For the elliptic Calogero-Moser problem related to the root system ${\bf B_2}$
the algebraic integrability with the explicit formulae for the additional
integrals were obtained in the recent paper~\cite{khodarinova:O}.
We use these formulae
to derive the spectral relations for the operator~(\ref{khodarinova:2}).

We would like to mention that although the procedure of the derivation of
the spectral relations provided the quantum integrals are given is
ef\/fective, the actual calculations are huge and would be very
dif\/f\/icult to
perform without computer. We have used a special program, which has been
created for this purpose.

\section{Spectral relations for the three-particle elliptic
Calogero-Moser problem}

Let's consider the quantum problem with the Hamiltonian~(\ref{khodarinova:1}). This is a
particular case of the three-particle elliptic Calogero-Moser problem
corresponding to a special value of the parameter in the interaction. The
algebraic integrability in this case has been conjectured by Chalykh and
Veselov in~\cite{khodarinova:ChV} and proved later
in~\cite{khodarinova:SV,khodarinova:Kh} and (in much
more general case) in~\cite{khodarinova:BEG}. We should mention that
only the paper~\cite{khodarinova:Kh}
contains the explicit formulae for the additional integrals, the other
proofs are indirect.

The usual integrability of this problem has been established by Calogero,
Marchioro and Ragnisco in~\cite{khodarinova:CMR}. The corresponding
integrals have the
form
\[
\ba{l}
L_1=L=-\Delta +4(\wp _{12}+\wp _{23}+\wp _{31}), \vspace{1mm}\\
L_2=\partial _1+\partial _2+\partial _3,
\vspace{1mm}\\
L_3=\partial _1\partial _2\partial _3+2\wp _{12}\partial _3+2\wp
_{23}\partial _1+2\wp _{31}\partial _2,
\ea
\]
where we have used the notations $\partial _i={\partial }/{\partial x_i}$,
$\wp _{ij}=\wp (x_i-x_j)$.

The following additional integrals have been found in~\cite{khodarinova:Kh}:
\[
\ba{l}
\ds I_{12}=(\partial _1-\partial _3)^2(\partial _2-\partial _3)^2-8\wp
_{23}(\partial _1-\partial _3)^2-8\wp _{13}(\partial _2-\partial _3)^2
\vspace{1mm}\\
\qquad {}+4(\wp _{12}-\wp _{13}-\wp _{23})(\partial _1-\partial _3)(\partial
_2-\partial _3) -2(\wp _{12}^{\prime }+\wp _{13}^{\prime }+6\wp _{23}^{\prime
})(\partial _1-\partial _3)
\vspace{1mm}\\
\qquad {}-2(-\wp _{12}^{\prime }+6\wp _{13}^{\prime }+\wp _{23}^{\prime
})(\partial _2-\partial _3)
-2\wp _{12}^{\prime \prime }-6\wp _{13}^{\prime \prime }-6\wp
_{23}^{\prime \prime }+4(\wp _{12}^2+\wp _{13}^2+\wp _{23}^2)
\vspace{1mm}\\
\qquad {}+8(\wp _{12}\wp _{13}+\wp _{12}\wp _{23}+7\wp _{13}\wp _{23}),
\ea
\]
two other integrals $I_{23}$, $I_{31}$ can be written simply by permuting
the indices. Unfortunately, none of these operators implies algebraic
integrability, because the symbols do not take dif\/ferent values on the
solutions of the system $P_i(\xi )=c_i$, $i=1,2,3$ (see the def\/inition in the
Introduction). But any non-symmetric linear combination of them, e.g.
$L_4=I_{12}+2I_{23}$ would f\/it into the def\/inition.

\medskip

\noindent
{\bf Lemma.} {\it The operators $L_1$, $L_2$, $L_3$, $I$,  where $I$  is
equal to $I_{12}$, $I_{23}$ or $I_{13}$,  satisfy the algebraic
relation:
\begin{equation}  \label{khodarinova:3}
Q(L_1,L_2,L_3,I) = I^3 + A_1I^2 + A_2I + A_3 = 0,
\end{equation}
where
\[
\ba{l}
 A_1 = 6g_2-X^2, A_2 = 2XY-15g_2^2-2g_2X^2,
\vspace{1mm}\\
A_3 = -Y^2-2g_2XY-108g_3Y+16g_3X^3+15g_2^2X^2-100g_2^3,
\vspace{1mm}\\
 X = 3/2L_1 + 1/2L_2^2,
\vspace{1mm}\\
 Y = 1/2L_1^3 + 27L_3^2 + 1/4L_2^6 + L_1L_2^4 - 5L_2^3L_3 +
5/4L_1^2L_2^2 - 9L_1L_2L_3.
\ea
\]}

The idea of the proof is the following. Let the relation $Q$ be a polynomial
of third degree of $I$ such that $I_{12}$, $I_{23}$, $I_{13}$ are its roots:
\[
Q = I^3 + A_1I^2 + A_2I + A_3.
\]
Then $A_1 = -(I_{12}+I_{23}+I_{13})$, $A_2 = I_{12}I_{23} + I_{12}I_{13} +
I_{23}I_{13}$, $A_3 = -I_{12}I_{23}I_{13}$. From the explicit formulae for
$I_{ij}$ it follows that the operators $A_i$ are symmetric and their highest
symbols $a_i$ are constants. So there exist polynomials $p_i$ such that $a_i
= p_i(l_1,l_2,l_3)$. Consider $A^{\prime}_i = A_i - p_i(L_1,L_2,L_3)$,
$A^{\prime}_i$ commute with $L$. It follows from the Berezin's lemma~\cite{khodarinova:B}
that if a dif\/ferential operator commutes with a Schr\"odinger operator then
the coef\/f\/icients in the highest symbol of this operator are polynomials in $x
$. Since the coef\/f\/icients in the highest symbols of the operators
$A^{\prime}_i$ are some elliptic functions, they must be constant in $x$. It
is clear that the operators $A^{\prime}_i$ are also symmetric and $\deg
a^{\prime}_i < \deg a_i$. So we can continue this procedure until we come to
zero. Thus we express $A_i$ as the polynomials of $L_1$, $L_2$ and $L_3$. To
calculate the explicit expressions we can use the fact that the coef\/f\/icients
in the highest symbols $a^{\prime}_i$ are constant and therefore may be
calculated at some special point. The most suitable choice is when
$x_1-x_2=\omega_1$, $x_2-x_3=\omega_2$, $x_1-x_3=(\omega_1+\omega_2)$,
where $2\omega_1$, $2\omega_2$ are the periods of Weierstrass $\wp$-function.
Then $\wp^{\prime}(x_1-x_2)=\wp^{\prime}(x_2-x_3)=\wp^{\prime}(x_1-x_3)=0$ and
$\wp(x_1-x_2)=e_1$, $\wp(x_2-x_3)=e_2$, $\wp(x_1-x_3)=e_3$, where $e_1$, $e_2$,
$e_3$ are the roots of the polynomial $4z^3 - g_2 z - g_3 = 0$ (see e.g.~\cite{khodarinova:WW})

\medskip

\noindent
{\bf Theorem 1.} {\it The integrals of the three-particle elliptic
Calogero-Moser problem~(\ref{khodarinova:1}) $L_1$, $L_2$,  $L_3$, $I = I_{12}$, $J =
I_{23}$ satisfy the algebraic system:
\[
\ba{l}
I^3 + A_1I^2 + A_2I + A_3 = 0,
\vspace{1mm}\\
I^2 + IJ + J^2 + A_1(I+J) + A_2 = 0,
\ea
\]
where $A_1$, $A_2$,  $A_3$  are given by the
 formulae~(\ref{khodarinova:3}). An additional
integral $L_4$ which guarantees the algebraic integrability
 of the problem can be chosen as $L_4 = I+2J$.}

\medskip

\noindent
{\bf Proof.} The f\/irst relation for $I$ is proved in the lemma. The proof of
the second relation is the following. The operators $I_{12} = I$, $I_{23} = J
$, $I_{13}$ satisfy the relations: $I+J+I_{13} = -A_1$, $IJ + (I+J)I_{13} =
A_2$, so $I_{13} = -(I+J+A_1)$ and therefore we obtain $IJ - (I+J)(I+J+A_1)
= A_2$. This completes the proof of Theorem 1.

\medskip

\noindent
{\bf Remark.} Putting $L_2=0$ we can reduce the problem~(\ref{khodarinova:1}) to the
plane $x_1+x_2+x_3=0$ and obtain two-dimensional elliptic Calogero--Moser
problem related to the root system~${\bf A_2}$. The spectral curve of this
problem was found in the paper~\cite{khodarinova:SV}. The corresponding formula from~\cite{khodarinova:SV}
has the form:
\[
\nu^3 + (6\lambda\mu^2-3(\lambda^2-3g_2)^2)\nu
-\mu^4+(10\lambda^3-18g_2\lambda+108g_3)\mu^2 + 2(\lambda^2-3g_2)^3 = 0.
\]
If in our formula~(\ref{khodarinova:3}) put $L_2=0$, substitute $L_1=2\lambda$,
$L_3=\sqrt{3}/9\mu$, $I=\nu-1/3(6g_2-X^2)$, then for the relation~(\ref{khodarinova:3}) we get:
\[
\nu^3 + (6\lambda\mu^2-3(\lambda^2-3g_2)^2)\nu
-\mu^4+(10\lambda^3-18g_2\lambda-108g_3)\mu^2 + 2(\lambda^2-3g_2)^3 = 0.
\]
The elliptic curves $y^2 = 4x^3-g_2x-g_3$ and $y^2 = 4x^3-g_2x+g_3$ are
isomorphic: $x\to -x$, $y\to iy$, so the dif\/ference in the sign by $g_3$ is
not important.

\section{Spectral relations for the elliptic Calogero-Moser problem
related to the root system ${\bf B_2}$}

Consider now the Schr\"odinger operator~(\ref{khodarinova:2}). Its algebraic
integrability conjectured in~\cite{khodarinova:ChV} has been proved in~\cite{khodarinova:O}.

The formulae for the quantum integrals in this case are (see~\cite{khodarinova:O})
\[
\ba{l}
L_1=L=-\Delta +2(\wp (x)+\wp (y)+2\wp (x+y)+2\wp (x-y)), \vspace{2mm}\\
L_2=\partial _x^2\partial _y^2-2\wp (y)\partial _x^2-2\wp (x)\partial
_y^2-4(\wp (x+y)-\wp (x-y))\partial _x\partial _y
\vspace{1mm}\\
\qquad {}-2(\wp ^{\prime }(x+y)+\wp ^{\prime }(x-y))\partial _x
-2(\wp^{\prime }(x+y)-\wp ^{\prime }(x-y))\partial _y \vspace{1mm}\\
\qquad {}-2(\wp ^{\prime \prime }(x+y)+\wp ^{\prime \prime }(x-y))+4(\wp
^2(x+y)+\wp ^2(x-y)) \vspace{1mm}\\
\qquad {}+4(\wp (x)+\wp (y))(\wp (x+y)+\wp (x-y)) -8\wp (x+y)\wp (x-y)-4\wp (x)\wp (y),
\vspace{2mm}\\
L_3=I_x+2I_y,
\ea
\]
where
\[
\ba{l}
I_x=\partial _x^5-5\partial _x^3\partial _y^2-10(1/2\wp (x)-\wp (y)+\wp
(x+y)+\wp (x-y))\partial _x^3 \vspace{1mm}\\
\qquad {}+30(\wp (x+y)-\wp (x-y))\partial _x^2\partial _y+15\wp (x)\partial
_x\partial _y^2 \vspace{1mm}\\
\qquad {}-15/2\wp (x)(\partial _x^2-\partial _y^2)+30(\wp (x+y)-\wp
(x-y))\partial _x\partial _y
\vspace{1mm}\\
\qquad {}+(10\wp ^{\prime \prime }(x+y)-10\wp ^{\prime \prime }(x-y)-30\wp
(y)(\wp (x+y)-\wp (x-y)))\partial _y \vspace{1mm}\\
\qquad {}+(30\wp (y)(\wp (x)-\wp (x+y)-\wp (x-y))+120\wp (x+y)\wp (x-y)
\vspace{1mm}\\
\qquad {}+10\wp ^{\prime \prime }(x+y)+10\wp ^{\prime \prime }(x-y)-5\wp
^{\prime \prime }(x)-9/2g_2)\partial _x \vspace{1mm}\\
\qquad {}-15(\wp ^{\prime }(x+y)+\wp ^{\prime }(x-y))(\wp (x)+\wp (y))
-15(\wp ^{\prime }(x)\wp (y)+\wp ^{\prime }(y)\wp (x)) \vspace{1mm}\\
\qquad {}+60(\wp ^{\prime }(x+y)+\wp ^{\prime }(x-y))(\wp (x+y)+\wp (x-y)),
\ea
\]
operator $I_y$ can be written by exchanging in the previous formula $x$ and
$y$, and we use the notations $\partial _x={\partial }/{\partial x},\partial
_y={\partial }/{\partial y}$.

\medskip

\noindent
{\bf Theorem 2.} {\it The quantum integrals $L= 1/2 L_1$, $M= L_2$, $I = I_x$,
$J = I_y$, $L_3 = I+2J$  of the elliptic Calogero-Moser system related to
the root system ${\bf B_2}$  satisfy the following algebraic relations:
\[
I^4+B_1I^2+B_2=0, \qquad  I^2+J^2+B_1 = 0,
\]
where
\[
\ba{l}
B_1= 32L^5-120ML^3+120M^2L+g_2(-82L^3+114LM) \vspace{1mm}\\
\qquad{}+g_3(-270L^2+486M)+102g_2^2L+486g_3g_2, \\
 B_2=400M^3L^4-1440M^4L^2+1296M^5
\vspace{1mm}\\
\qquad{}+g_2(-400ML^6+840M^2L^4 -576M^3L^2+648M^4) \vspace{1mm}\\
\qquad{}+g_3(800L^7-8280ML^5+22032M^2L^3 -17496LM^3) \vspace{1mm}\\
\qquad{}+g_2^2(800L^6-1815ML^4+3510M^2L^2-3807M^3) \vspace{1mm}\\
\qquad{}+g_3g_2(3870L^5+324ML^3-13122M^2L) \vspace{1mm}\\
\qquad{}+g_3^2(18225L^4-65610ML^2+59049M^2) \vspace{1mm}\\
\qquad{}+g_2^3(-2930L^4+5418ML^2-4536M^2) \vspace{1mm}\\
\qquad{} +g_3g_2^2(-21708L^3+26244LM) \vspace{1mm}\\
\qquad{}+g_3^3g_2(-65610L^2+118098M)+g_2^4(2772L^2-1539M) \vspace{1mm}\\
\qquad{}+21870g_3g_2^3L+59049g_2^2g_3^2-162g_2^5.
\ea
\]}

The proof is analogous to the previous case. The f\/irst relation is a
polynomial of second order in $I^2$ such that $I_x^2$ and $I_y^2$ are its
roots. Calculations of the coef\/f\/icients as in the previous case can be done
at the special point $(x,y):$ $x = \omega_1$, $y = \omega_2$. Then
$\wp(x)=e_1$, $\wp(y)=e_2$, $\wp(x+y)=\wp(x-y)=e_3$,
$\wp^{\prime}(x)=\wp^{\prime}(y)=\wp^{\prime}(x-y)=\wp^{\prime}(x-y)=0$.

As we have mentioned already these calculations have been done with the help
of the special computer program created by the authors.

\subsection*{Acknowledgment}

\noindent
The authors are grateful to O.A.~Chalykh and  A.P.~Veselov for useful
discussions and to the referees for the helpful critical remarks.

 \label{khodarinova-lp}

\end{document}